\documentstyle[12pt]{article}
\def\thebibliography#1{\bigskip\section*{\centering
References\\}\bigskip\list
{\arabic{enumi}.}{\settowidth\labelwidth{#1}\leftmargin\labelwidth
\advance\leftmargin\labelsep
\usecounter{enumi}}
\def\newblock{\hskip .11em plus .33em minus .07em}
\sloppy\clubpenalty4000\widowpenalty4000
\sfcode`\.=1000\relax}

\def\op#1{\mathop{\fam0 #1}\limits}

\newcommand{\Id}{{\rm Id\,}}
\def\Ker{{\rm Ker\,}}
\newcommand{\ben}{\begin{eqnarray}}
\newcommand{\een}{\end{eqnarray}}
\newcommand{\be}{\begin{eqnarray*}}
\newcommand{\ee}{\end{eqnarray*}}
\newcommand{\bea}{\begin{eqalph}}
\newcommand{\eea}{\end{eqalph}}
\newcommand{\cL}{{\cal L}}
\newcommand{\cE}{{\cal E}}
\newcommand{\cH}{{\cal H}}
\newcommand{\cF}{{\cal F}}
\newcommand{\cG}{{\cal G}}
\newcommand{\al}{\alpha}
\newcommand{\bt}{\beta}
\newcommand{\dl}{\delta}
\newcommand{\la}{\lambda}
\newcommand{\La}{\Lambda}

\newcommand{\om}{\omega}
\newcommand{\Om}{\Omega}
\newcommand{\m}{\mu}

\newcommand{\g}{\gamma}
\newcommand{\G}{\Gamma}

\newcommand{\ve}{\varepsilon}
\newcommand{\th}{\theta}

\newcommand{\w}{\wedge}

\newcommand{\wt}{\widetilde}
\newcommand{\wh}{\widehat}
\newcommand{\ol}{\overline}
\newcommand{\dr}{\partial}

\newcounter{eqalph}
\newcounter{equationa}

\newenvironment{eqalph}{\stepcounter{equation}
\setcounter{equationa}{\value{equation}}
\setcounter{equation}{0}

\begin{eqnarray}}{\end{eqnarray}
\setcounter{equation}{\value{equationa}}}

\begin{document}
\hbox{}

\centerline{\bf\large ENERGY-MOMENTUM CONSERVATION LAW}
\medskip

\centerline{\bf\large IN HAMILTONIAN FIELD THEORY}
\bigskip

\centerline{\bf Gennadi A Sardanashvily}
\medskip

\centerline{Department of Theoretical Physics, Physics Faculty,}

\centerline{Moscow State University, 117234 Moscow, Russia}

\centerline{E-mail: sard@grav.phys.msu.su}
\bigskip

\begin{abstract}
In the Lagrangian field theory, one gets different identities for different
stress energy-momentum tensors, e.g., canonical energy-momentum
tensors. Moreover, these identities are not conservation laws of the
above-mentioned energy-momentum tensors in general. In the framework of the
multimomentum Hamiltonian formalism, we have the fundamental identity whose
restriction to a constraint space can be treated the energy-momentum
conservation law. In standard field models, this appears the metric
energy-momentum conservation law.
\end{abstract}

\section{Introduction}

We follow the generally accepted geometric description of classical
fields by sections of fibred manifolds $Y\to X.$
Their dynamics is phrased in terms of jet spaces (see \cite{got,kol,sard,8sar}
for the bibliography). Given a fibred manifold
$Y\to X$, the $k$-order jet space $J^kY$ of $Y$
comprises the equivalence classes
$j^k_xs$, $x\in X$, of sections $s$ of $Y$ identified by the first $(k+1)$
terms of their Taylor series at a point $x$.
One exploits
the well-known facts
that: (i) the $k$-jet space of sections of a fibred
manifold $Y$ is a finite-dimensional smooth manifold and (ii)
a $k$-order differential operator on sections of a fibred manifold $Y$
can be described as a morphism of $J^kY$ to a vector bundle over $X$.
As a consequence, the dynamics of field systems is played out
on finite-dimensional configuration and phase spaces. Moreover, this
dynamics is phrased in the geometric terms due to
the 1:1 correspondence between the sections of the jet bundle
$J^1Y\to Y$ and the connections on the fibred manifold $Y\to X$
\cite{sard,8sar,sau}.

In field theory, we can restrict ourselves
to the first order  Lagrangian formalism when the configuration space
is $J^1Y$. Given fibred coordinates $(x^\m, y^i)$
of $Y$, the jet space $J^1Y$ is endowed with the adapted  coordinates
$ (x^\m, y^i, y^i_\m)$.
A first order Lagrangian density on the configuration space $J^1Y$ is
represented by a horizontal exterior density
\[L=\cL(x^\m, y^i, y^i_\m)\om, \qquad \om=dx^1\w ...\w dx^n,
\qquad n=\dim X.\]
The corresponding first order Euler-Lagrange equations for sections
$\ol s$ of the fibred jet manifold $J^1Y\to X$ read
\ben &&\dr_\la\ol s^i=\ol s^i_\la, \nonumber\\
&& \dr_i\cL-(\dr_\la+\ol s^j_\la\dr_j
+\dr_\la\ol s^j_\m\dr^\m_j)\dr^\la_i\cL=0. \label{306}\een

Conservation laws are usually related with symmetries.
There are several approaches to examine symmetries in the Lagrangian
formalism, in particular, in jet terms \cite{got2,gri,8sar}. We consider the
Lie derivatives of Lagrangian densities in order to obtain differential
conservation laws. These are conservation laws of Noether currents and
the energy-momentum conservation laws.

If the Lie derivative of a
Lagrangian density $L$ along a vertical vector field $u_\cG$ on a bundle
$Y$ is equal to zero, the current conservation law on solutions of the
first order Euler-Lagrange equations (\ref{306}) takes place. In this case,
the vertical vector field $u_\cG$ plays the role of a generator of
internal symmetries.

In case of the energy-momentum conservation law, a vector field on $Y$ is
not vertical and the Lie derivative of $L$ does not vanish in general.
Therefore, this conservation law is not related with a symmetry.
Moreover, one can not say {\it a priori} what is conserved.

In gravitation theory, the first integral of gravitational equations is
the conservation law of the metric energy-momentum tensor of matter, but only
in the presence of the gravitational field generated by this matter.
In other models, the metric energy-momentum tensor holds {\it a posteriori}.

Let
\[ u=u^\mu\dr_\mu + u^i\dr_i\]
be a vector field on a fibred manifold $Y$ and $\ol u$ its jet lift
onto the fibred jet manifold $J^1Y\to X$. Given a
Lagrangian density $L$ on $J^1Y$, let us computer the Lie derivative
${\bf L}_{\ol u}L$. On solutions $\ol s$ of the first order Euler-Lagrange
equations (\ref{306}), we have the equality
\begin{equation}
\ol s^*{\bf L}_{\ol u}L= \frac{d}{dx^\la}[\pi^\la_i(\ol s)(u^i-u^\mu \ol
s^i_\mu) +u^\la\cL (\ol s)]\om, \qquad \pi^\mu_i=\dr^\mu_i\cL. \label{502}
\end{equation}
In particular, if $u$ is a vertical vector field such that
\[{\bf L}_{\ol u}L=0,\]
the equality (\ref{502}) takes the form of the current
conservation law
\begin{equation}
\frac{d}{dx^\la}[u^i\pi^\la_i(\ol s)]=0. \label{503}
\end{equation}
In gauge theory, this conservation law is exemplified by the Noether
identities.

Let
\[\tau=\tau^\la\dr_\la\]
be a vector field on $X$ and
\[ u=\tau_\G=\tau^\mu (\dr_\mu+\G^i_\mu\dr_i)\]
its horizontal lift onto the fibred manifold $Y$ by a connection $\G$
on $Y$. In this case, the equality (\ref{502}) takes the form
\begin{equation}
\ol s^*{\bf L}_{\ol\tau_\G}L=
-\frac{d}{dx^\la}[\tau^\mu T_\G{}^\la{}_\mu (\ol s)]\om \label{504}
\end{equation}
where
\begin{equation}
T_\G{}^\la{}_\mu (\ol s) =\pi^\la_i(\ol s^i_\mu-\G^i_\mu)
-\delta^\la_\mu\cL \label{84}
\end{equation}
is the canonical energy-momentum
tensor of a field $\ol s$ with respect to the connection $\G$ on $Y$.
The tensor (\ref{84}) is the particular case of the
stress energy-momentum tensors \cite{fer,got2,kij}.

Note that, in
comparison with the current conservation laws, the Lie derivative
${\bf L}_{\ol\tau_\G}L$ fails to be equal to zero as a rule, and the equality
(\ref{504}) is not the conservation law of any canonical
energy-momentum tensor in general. It does not look fundamental, otherwise
its Hamiltonian counterpart.
In the framerwork of the multimomentum Hamiltonian formalism, we get the
fundamental identity (\ref{5.27}) whose restriction to a constraint
space can be treated the energy-momentum
conservation law. In standard field models, this appears the metric
energy-momentum conservation law in the presence of a background world metric.

Lagrangian densities of field models are almost always degenerate
and the corresponding Euler-Lagrange
equations are underdetermined. To describe constraint field systems,
the multimomentum Hamiltonian formalism can be utilized
\cite{6sar,7sar,lsar}.
In the framework of this formalism,
the phinite-dimensional phase space of fields is the Legendre bundle
\begin{equation}
\Pi=\op\w^n T^*X\op\otimes_Y TX\op\otimes_Y V^*Y \label{00}
\end{equation}
over $Y$ into which the Legendre
morphism $\wh L$ associated with a Lagrangian density
$L$ on $J^1Y$ takes its values. This phase space is
provided with  the fibred coordinates $(x^\la ,y^i,p^\la_i)$ such that
\[ (x^\m,y^i,p^\m_i)\circ\wh L=(x^\m,y^i,\pi^\m_i).\]
The Legendre bundle (\ref{00}) carries the multisymplectic form
\begin{equation}
\Om =dp^\la_i\w
dy^i\w\om\otimes\dr_\la. \label{406}
\end{equation}
In case of $X={\bf R}$, the form $\Om$ recovers
the standard symplectic form in analytical mechanics.

Building on the multisymplectic form $\Om$, one can develop
the so-called multimomentum Hamiltonian formalism of field theory where
canonical momenta correspond to derivatives of fields with respect to all world
coordinates, not only the temporal one. On the mathematical level, this is
the straightforward multisymplectic
generalization of the standard Hamiltonian formalism in analytical mechanics
to fibred manifolds over an $n$-dimensional base $X$, not only ${\bf R}$.

Note that the Hamiltonian approach to field theory was called into
play mainly for canonical quantization of fields by analogy with quantum
mechanics. The major goal of this approach has consisted in establishing
simultaneous commutation relations of quantum fields in models with
degenerate Lagrangian densities, e.g., gauge theories.
In classical field theory, the conventional Hamiltonian formalism
fails to be so successful. In the straightforward manner, it takes the form
of the instantaneous Hamiltonian formalism when canonical variables are
field functions at a given instant  of time. The corresponding phase space
is infinite-dimensional. Hamiltonian dynamics played out on this phase
space is far from to be a partner
of the usual Lagrangian dynamics of
field systems. In particular, there are no Hamilton equations in the
bracket form which would be adequate to Euler-Lagrange field
equations, otherwise in the multumomentum Hamiltonian formalism.

We say that a connection $\g$ on the fibred Legendre manifold $\Pi\to
X$ is a Hamiltonian connection if the form $\g\rfloor\Om$ is closed.
Then, a Hamiltonian form $H$ on $\Pi$ is defined to be an
exterior form such that
\begin{equation}
dH=\g\rfloor\Om \label{013}
\end{equation}
for some Hamiltonian connection $\g$. Every  Hamiltonian form admits splitting
\begin{equation}
H =p^\la_idy^i\w\om_\la -p^\la_i\G^i_\la\om
-\wt{\cH}_\G\om=p^\la_idy^i\w\om_\la-\cH\om,\qquad
\om_\la=\dr_\la\rfloor\om,  \label{017}
\end{equation}
where $\G$ is a connection on $Y\to X$.
Given the  Hamiltonian form $H$ (\ref{017}), the equality
(\ref{013}) comes to the Hamilton equations
\begin{equation}
\dr_\la r^i(x) =\dr^i_\la\cH, \qquad
\dr_\la r^\la_i(x)=-\dr_i\cH \label{3.11}
\end{equation}
for sections $r$ of the fibred Legendre manifold $\Pi\to X$.

The Hamilton equations (\ref{3.11}) are the multimomentum generalization
of the standard Hamilton equations in mechanics when $X={\bf R}$.
The energy-momentum conservation law (\ref{5.27}) which we suggest is
accordingly
the multimomentum generalization of the familiar energy conservation law
\begin{equation}
\frac{d\cH}{dt}=\frac{\dr\cH}{\dr t} \label{E1}
\end{equation}
in analytical mechanics.

\section{Technical preliminary}

A fibred manifold
\[\pi:Y\to X\]
is provided with fibred coordinates $(x^\la, y^i)$ where $x^\la$ are
coordinates of the base $X$.
A locally trivial fibred manifold is termed the bundle.
We denote by $VY$ and $V^*Y$ the vertical tangent bundle and the
vertical cotangent bundle of $Y$ respectively.
For the sake of simplicity, the pullbacks
$Y\op\times_XTX$ and $Y\op\times_XT^*X$
are denoted by $TX$ and $T^*X$ respectively.

On fibred manifolds, we
consider  the following types of differential forms:

(i) exterior horizontal forms $ Y\to\op\w^r T^*X$,

(ii) tangent-valued horizontal forms $Y\to\op\w^r T^*X\op\otimes_Y TY$
and, in particular, soldering forms $Y\to T^*X\op\otimes_YVY$,

(iii) pullback-valued forms
\be &&Y\to \op\w^r T^*Y\op\otimes_Y TX, \\
&&Y\to \op\w^r T^*Y\op\otimes_Y T^*X. \ee
Horizontal $n$-forms are called horizontal densities.

Given a fibred manifold $Y\to X$, the first order jet manifold $J^1Y$ of
$Y$ is both the fibred manifold $J^1Y\to X$
and the affine bundle $J^1Y\to Y $  modelled on the vector
bundle $T^*X\otimes_Y VY.$
It is endowed with the adapted coordinates $(x^\la,y^i,y^i_\la)$:
\[{y'}^i_\la=(\frac{\dr {y'}^i}{\dr y^j}y_\m^j +
\frac{\dr{y'}^i}{\dr x^\m})\frac{\dr x^\m}{\dr{x'}^\la}.\]
We identify $J^1Y$ to its image under the  canonical bundle monomorphism
\ben &&\la:J^1Y\op\to_YT^*X \op\otimes_Y TY,\nonumber\\
&&\la=dx^\la\otimes(\dr_\la+y^i_\la \dr_i).\label{18}\een
Given a fibred morphism of $\Phi: Y\to Y'$
over a diffeomorphism of $X$, its jet prolongation
$J^1\Phi:J^1Y\to J^1Y'$ reads
\[ {y'}^i_\mu\circ
J^1\Phi=(\dr_\la\Phi^i+\dr_j\Phi^iy^j_\la)\frac{\dr x^\la}{\dr {x'}^\mu}.\]

Every  vector field
\[u = u^\la\dr_\la + u^i\dr_i\]
on a fibred manifold $Y\to X$ gives rise to the projectable vector field
\ben &&\overline u =r_1\circ J^1u: J^1Y\to J^1TY\to TJ^1Y,\nonumber\\
&& \overline u =
u^\la\dr_\la + u^i\dr_i + (\dr_\la u^i+y^j_\la\dr_ju^i
- y_\m^i\dr_\la u^\m)\dr_i^\la, \label{1.21}\een
on the fibred jet manifold $J^1Y\to X$ where $J^1TY$ is the jet
manifold of the fibred manifold $TY\to X$.

The canonical morphism (\ref{18}) gives rise to the bundle
monomorphism
\[
 \wh\la: J^1Y\op\times_X TX\ni\dr_\la\mapsto\wh{\dr}_\la = \dr_\la\rfloor
\la\in J^1Y\op\times_Y TY,
\]
\[
\wh{\dr}_\la=\dr_\la + y^i_\la \dr_i.
\]
This morphism determines the canonical horizontal
splitting of the pullback
\begin{equation}
J^1Y\op\times_Y TY=\wh\la(TX)\op\oplus_{J^1Y} VY,\label{1.20}
\end{equation}
\[
\dot x^\la\dr_\la
+\dot y^i\dr_i =\dot x^\la(\dr_\la +y^i_\la\dr_i) + (\dot y^i-\dot x^\la
y^i_\la)\dr_i.
\]
In other words, over $J^1Y$, we have the canonical horizontal splitting of
the tangent bundle $TY$.

Building on the canonical splitting (\ref{1.20}),
one gets the corresponding horizontal splittings of
a projectable vector field
\begin{equation}
u =u^\la\dr_\la +u^i\dr_i=u_H +u_V =u^\la (\dr_\la +y^i_\la
\dr_i)+(u^i - u^\la y^i_\la)\dr_i \label{31}
\end{equation}
on a fibred manifold $Y\to X$.

Given a fibred  manifold  $Y\to X$, there is
the 1:1 correspondence between the connections on $Y\to X$
and global sections
\[\G =dx^\la\otimes(\dr_\la+\G^i_\la\dr_i)\]
of the affine jet bundle $J^1Y\to Y$. Substitution of such a global
section $\G$ into the canonical horizontal splitting (\ref{1.20})
recovers the familiar horizontal splitting of the tangent bundle $TY$
with respect to the connection $\G$ on $Y$. These global sections form the
affine space modelled on the linear space of soldering forms on $Y$.

Every connection $\G$ on $Y\to X$ yields the first order differential
operator
\be && D_\G:J^1Y\op\to_YT^*X\op\otimes_YVY,\\
&&D_\G=(y^i_\la-\G^i_\la)dx^\la\otimes\dr_i,\ee
on $Y$ which is called the covariant differential relative to the
connection $\G$.

The repeated jet manifold
$J^1J^1Y$, by definition, is the first order jet manifold of
$J^1Y\to X$. It is provided with the adapted coordinates
$(x^\la ,y^i,y^i_\la ,y_{(\m)}^i,y^i_{\la\m})$.
Its subbundle $ \wh J^2Y$ with $y^i_{(\la)}= y^i_\la$ is called the
sesquiholonomic jet manifold.
The second order jet manifold $J^2Y$ of $Y$ is the subbundle
of $\wh J^2Y$ with $ y^i_{\la\m}= y^i_{\m\la}.$

There exists the following generalizations of the contact map
(\ref{18}) to the second order jet manifold $J^2Y$:
\ben
&&\la:J^2Y\op\to_{J^1Y}
T^*X \op\otimes_{J^1Y} TJ^1Y,\nonumber\\
 &&\la=dx^\la\otimes\wh{\dr}_\la=dx^\la
\otimes (\dr_\la + y^i_\la \dr_i + y^i_{\m\la}\dr_i^\m),\label{54}
\een
The contact map (\ref{54}) defines the canonical horizontal splitting of the
tangent bundle $TJ^1Y$ and the corresponding horizontal splitting of
a projectable vector field $\ol u$ on $J^1Y$ over $J^2Y$:
\ben
&&\ol u=u_H+u_V = u^\la(\dr_\la+ y^i_\la\dr_i+y^i_{\m\la}\dr^\mu_i)
\nonumber \\
&& \qquad + [(u^i-y^i_\la u^\la)\dr_i + (u^i_\m- y^i_{\m\la}u^\la)\dr^\m_i] .
\label{79}
\een

\section{Conservation laws in the Lagrangian formalism}

Let $Y\to X$ be a fibred manifold and $ L=\cL\om$ a Lagrangian density
on $J^1Y$.  With $L$, the jet manifold $J^1Y$  carries
the uniqie associated Poincar\'e-Cartan form
\begin{equation}
\Xi_L=\pi^\la_idy^i\w\om_\la -\pi^\la_iy^i_\la\om +\cL\om \label{303}
\end{equation}
and the Lagrangian  multisymplectic form
\[\Om_L=(\dr_j\pi^\la_idy^j+\dr^\m_j\pi^\la_idy^j_\m)\w
dy^i\w\om\otimes\dr_\la.\]
Using the pullback of these forms onto the
repeated jet manifold $J^1J^1Y$, one can construct the exterior form
\begin{equation}
\La_L=d\Xi_L-\la\rfloor\Om_L=[y^i_{(\la)}-y^i_\la)d\pi^\la_i +
(\dr_i-\wh\dr_\la\dr^\la_i)\cL dy^i]\w\om,\label{304}
\end{equation}
\[ \la=dx^\la\otimes\wh\dr_\la,\qquad
\wh\dr_\la =\dr_\la +y^i_{(\la)}\dr_i+y^i_{\m\la}\dr^\m_i,\]
on $J^1J^1Y$.
Its restriction to the second order jet manifold $J^2Y$ of $Y$ reproduces
the familiar variational Euler-Lagrange operator
\begin{equation}
\cE_L= [\dr_i-
(\dr_\la +y^i_\la\dr_i+y^i_{\m\la}\dr^\m_i)\dr^\la_i]\cL dy^i\w\om.\label{305}
\end{equation}
The restriction of the form (\ref{304}) to the sesquiholonomic jet manifold
$\wh J^2Y$ defines the sesquiholonomic extension
$\cE'_L$  of the Euler-Lagrange operator (\ref{305}).
It is given by the expression
(\ref{305}), but with nonsymmetric coordinates $y^i_{\m\la}$.

Let $\ol s$ be a section of the fibred jet manifold $J^1Y\to X$ such that
its first order jet prolongation  $J^1\ol s$ takes its values into
$\Ker\cE'_L$. Then, $\ol s$ satisfies the first order
differential Euler-Lagrange equations (\ref{306}).
They are equivalent to the second order Euler-Lagrange equations
\begin{equation}
\dr_i\cL-(\dr_\la+\dr_\la s^j\dr_j
+\dr_\la\dr_\mu s^j \dr^\m_j)\dr^\la_i\cL=0.\label{2.29}
\end{equation}
for sections $s$ of $Y$ where $\ol s=J^1s$.

We have the following conservation laws on solutions of the first order
Euler-Lagrange equations.

Let
\[ u=u^\mu\dr_\mu + u^i\dr_i\]
be a vector field on a fibred manifold $Y$ and $\ol u$ its jet lift
(\ref{1.21}) onto the fibred jet manifold $J^1Y\to X$. Given a
Lagrangian density $L$ on $J^1Y$, let us compute the Lie derivative
${\bf L}_{\ol u}L$. We have
\begin{equation}
{\bf L}_{\ol u}L= [\wh \dr_\la(\pi^\la_i(u^i-u^\mu y^i_\mu ) +u^\la\cL
)+ (u^i-u^\mu y^i_\mu )(\dr_i-\wh\dr_\la\dr^\la_i)\cL]\om, \label{501}
\end{equation}
\[\wh\dr_\la =\dr_\la +y^i_\la\dr_i+y^i_{\m\la}\dr^\m_i.\]
On solutions $\ol s$ of the first order Euler-Lagrange equations, the
equality (\ref{501}) comes to the conservation law (\ref{502}).

To calculate the Lie derivative (\ref{501}), on should single out the
vertical component $\ol u_V$ of the vector field $\ol u$ (\ref{1.21})
on $J^1Y$. Its horizontal splitting (\ref{79}) over $J^2Y$ reads
\ben
&&\ol u_H = u^\la(\dr_\la+ y^i_\la+y^i_{\m\la}) =u^\la\wh\dr_\la, \\
&& \ol u_V= (u^i-y^i_\la u^\la)\dr_i + (\dr_\la u^i + y^j_\la\dr_j u^i
- y^i_\mu\dr_\la u^\mu - y^i_{\la\mu}u^\mu)\dr^\la_i. \label{80}
\een
Given the splitting (\ref{80}), we have
\begin{equation}
{\bf L}_{\ol u}L = ({\bf L}_{\ol u_V}\cL)\om + \wh\dr_\la(u^\la\cL)\om.
\label{81}
\end{equation}
In particular, if
\begin{equation}
({\bf L}_{\ol u_V}\cL)=0, \label{82}
\end{equation}
the Lie derivative (\ref{81}) comes to the total differential.

If $u$ is a vertical vetor field such that the relation (\ref{82}) holds,
we have the conservation  law (\ref{502}) of the current $u^i\pi^\mu_i$
associated with the vertical field $u$.

If the vector field $u$ is not vertical,
the equation (\ref{82}) where $\ol u_V$
is given by the expression (\ref{80}) is formal because of the factor
$y^i_{\la\mu}$. To overcome this difficulty, one can construct the lift
$u_\infty$
of a projectable vector field $u$ on the bundle $Y$ onto the infinite order
jet space $J^\infty Y$. Moreover, the vertical part of $u_\infty$ must
take a certain form in order that the equation (\ref{82}) makes sense.
We have
\ben
&& u_\infty = u^\la(\dr_\la + y^i_\la\dr_i +...) + j^\infty
u_\cG,\label{83} \\
&& j^\infty u_\cG = \op\sum_{k=0}^\infty \wh\dr_{\la_k}...
\wh\dr_{\la_1}u^i\dr^{\la_1...\la_k}_i, \nonumber
\een
where $u_\cG$ is a certain vertical vector field on $Y$. In gauge theory,
$u_\cG$ are principal vector fields on a principal bundle $P$ which are
associated with isomorphisms of $P$. Next Section covers this case.

Now, let
\[ u=\tau_\G=\tau^\mu (\dr_\mu+\G^i_\mu\dr_i)\]
be the horizontal lift of a vector field
\[\tau=\tau^\la\dr_\la\]
on $X$ onto the fibred manifold $Y$ by a connection $\G$
on $Y$. In this case, the equality (\ref{502}) takes the form (\ref{504})
where $T_\G{}^\la{}_\mu (\ol s)$ (\ref{84}) are coefficients of the
$T^*X$-valued form
\begin{equation}
T_\G(\ol s)=-(\G\rfloor\Xi_L)\circ\ol s =[\pi^\la_i(\ol s^i_\mu-\G^i_\mu)
-\delta^\la_\mu\cL]dx^\mu\otimes\om_\la \label{S14}
\end{equation}
on $X$. One can think on this form as being the
canonical energy-momentum
tensor of a field $\ol s$ with respect to the connection $\G$ on $Y$. In
particular, when the fibration $Y\to X$ is trivial,
one can choose the trivial connection $\G=\theta_X$.
In this case, the form (\ref{S14}) is precisely the standard canonical
energy-momentum tensor. If
\[{\bf L}_\tau\cL=0\]
for all vector fields $\tau$ on $X$, the conservation law (\ref{504})
comes to the familiar conservation law
\[\frac{d}{dx^\la} T^\la{}_\mu (\ol s)=0\]
of the canonical energy-momentum tensor. In general,
the Lie derivative
${\bf L}_{\ol\tau_\G}L$ is not equal to zero and the equality
(\ref{504}) is not the conservation law of the canonical
energy-momentum tensor.

Note that if the above-mentioned field $\tau$ on $X$ is associated with a
diffeomorphism of the manifold $X$, its horizontal lift $\tau_\G$
corresponds to the lift of this diffeomorphism to the bundle
isomorphism of $Y$ by the connection $\G$ on $Y$. But the
equality (\ref{504}) fails to be a symmetry condition if the
corresponding Lie derivative is not equal to zero.

\section{Gauge symmetries}

In gauge theory, several types of gauge transformations are considered.
We follow the definition of gauge transformations as isomorphisms of a
principal bundle \cite{mar,soc}.

Let
\[
\pi_P :P\to X
\]
be a principal bundle with a structure
Lie group $G$ which acts freely and transitively on $P$ on the right:
\ben
&&r_g : p\mapsto pg, \label{1}\\
&& p\in P,\quad g\in G. \nonumber
\een
A principal bundle $P$ is also the general affine bundle modelled on the
left on the associated group bundle $\wt P$ with the standard fibre
$G$ on which the structure group $G$ acts by the adjoint representation.
The corresponding bundle morphism reads
\[
\wt P\times P\ni (\wt p,p)\mapsto \wt pp\in P.
\]
Note that the standard fibre of the group bundle $\wt P$ is the group
$G$ , while that of the principal bundle $P$ is the group
space of $G$ on which the structure group $G$ acts on the left.

A principal connection $A$ on
a principal bundle $P\to X$ is defined to be a
$G$-equivariant connection on $P$ such that
\[
J^1r_g\circ A= A\circ r_g
\]
for each canonical morphism (\ref{1}).
There is the 1:1 correspondence between the principal connections on a
principal bundle $P\to X$  and the global sections of the quotient
\begin{equation}
C=J^1 P/G\label{68}
\end{equation}
of the jet bundle $J^1 P\to P$ by the first
order jet prolongations of the canonical morphisms (\ref{1}).
We shall call
$C$ the principal connection bundle. It is an affine bundle modelled on
the vector bundle
\[
\ol C =T^*X \otimes V^GP
\]
where
\[
 V^GP=VP/G
\]
is the quotient of the vertical tangent bundle $VP$ of $P$
by the canonical action (\ref{1}) of $G$ on $P$.
Its standard fibre is the right
Lie algebra $\cG_r$ of the right-invariant vector fields on the group
$G$. The group $G$ acts on this standard fibre by the adjoint representation.

Given a bundle atlas $\Psi^P$ of $P$, the bundle $C$
is provided with  the fibred coordinates $(x^\mu,k^m_\mu)$ so that
\[(k^m_\mu\circ A)(x)=A^m_\mu(x)\]
are coefficients of the local connection 1-form of a principal connection
$A$ with respect to the atlas $\Psi^P$.
The first order jet manifold $J^1C$ of the bundle $C$ is
provided with the adapted coordinates $
(x^\mu, k^m_\mu, k^m_{\mu\la}).$

Let $Y\to X$ be a bundle associated with a principal bundle $P\to X$.
The structure group $G$ of $P$ acts freely on the standard fibre $V$ of
$Y$ on the left. The total space of the $P$-associated bundle $Y$,
by definition, is the quotient
\[
Y=(P\times V)/G
\]
of the product $P\times V$
by identification of its elements $(pg\times gv)$ for all $g\in G$.

Every principal connection $A$ on a principal bundle $P$ yields the
associated connection
\begin{equation}
\G=dx^\la\otimes [\dr_\la +A^m_\mu (x)I_m{}^i{}_jy^j\dr_i] \label{S4}
\end{equation}
where $A^m_\mu (x)$ are coefficients of the local connection 1-form
and $I_m$ are generators of the structure group $G$
on the standard fibre $V$ of the bundle $Y$.

Since only Lie derivatives along vertical vector fields lead to
conservation laws, we here restrict our consideration to isomorphisms
of a principal bundle $P$ over the identity morphism of its base $X$.
Given a principal bundle $P\to X$,
by a principal morphism is meant its $G$-equivariant
isomorphism $\Phi_P$ over $X$ together with the first order
jet prolongations $J^1\Phi_P$. Whenever $g\in G$, we have
\[
r_g\circ\Phi_P=\Phi_P\circ r_g.
\]
Every such isomorphism $\Phi_P$ is brought into the form
\begin{equation}
\Phi_P(p)=pf_s(p),  \qquad p\in P,\label{S15}
\end{equation}
where $f_s$ is a $G$-valued equivariant function on $P$:
\[
 f_s(qg)=g^{-1}f_s(q)g, \qquad g\in G.
\]
There is the 1:1 correspondense between these functions and the global
section $s$ of the group bundle $\wt P$:
\[
s(\pi(p))p=pf_s(p).
\]

For each $P$-associated
bundle $Y$, there exists the fibre-preserving representation morphism
\[
\wt P\times Y\ni (\wt p,y)\mapsto \wt py\in Y
\]
where $\wt P$ is the $P$-associated group bundle. Building on this
representation morphism, one can induce principal morphisms of $Y$:
\[
\Phi_s: Y\ni y\mapsto (s\circ \pi)(y)y\in Y
\]
where $s$ is a global section of $\wt P$.

Principal morphisms $\Phi_P$
constitute the gauge group which is isomorphic to the group of global
sections of the $P$-associated group bundle $\wt P$.
The Sobolev completion of the
gauge group is a Banach Lie group. Its Lie algebra
in turn is the Sobolev completion of the algebra of generators of
infinitesimal principal morphsms. These generators are represented by
the corresponding vertical vector fields $u_\cG$ on a $P$-associated
bundle $Y$ which
carries representation of the gauge group. We call them principal
vector fields.

It is readily observed that a
Lagrangian density $L$  on the configuration space $J^1Y$ is gauge
invariant iff, whenever local principal vector field $u_\cG$ on $Y\to X$,
\begin{equation}
{\bf L}_{\ol u_\cG} L=0.\label{85}
\end{equation}

In case
of unbroken symmetries, the total configuration space of gauge theory
is the product
\[J^1Y\op\times_X J^1C\]
where $Y$ is a $P$-associated vector bundle whose sections describe
matter fields.

Local principal vector fields on the $P$-associated vector bundle
$Y\to X$ read
\[
u_\cG=\al^m(x)I_m{}^i{}_jy^j\dr_i
\]
where $I_m$ are generators of the structure group $G$ acting on $V$
and $\al^m(x)$ are arbitrary local functions on $X$.
Local principal vector fields on the principal connection
bundle $C$ are written
\[
u_\cG=(\dr_\mu\al^m+c^m_{nl}k^l_\mu\al^n)\dr^\mu_m .
\]
Then, a local principal vector field on the product $C\op\times_XY$ takes
the form
\[
u_\cG = (u^A_m\al^m + u^{A\la}_m\dr_\la\al^m)\dr_A=
 (\dr_\mu\al^m+c^m_{nl}k^l_\mu\al^n)\dr^\mu_m +
\al^m(x)I_m{}^i{}_jy^j\dr_i
\]
where the collective index $A$ is utilized.
Substituting this expression into the equality
(\ref{85}), one recovers the familiar Noether identities for a gauge
invariant Lagrangian density $L$:
\be
&& u^A_m(\dr_A-\wh\dr_\la\dr^\la_A)\cL + \wh\dr_\la(u^A_m\dr^\la_A\cL)=0,\\
&& u^{A\mu}_m(\dr_A-\wh\dr_\la\dr^\la_A)\cL
+ \wh\dr_\la(u^{A\mu}_m\dr^\la_A\cL) + u^A_m\dr^\mu_A\cL =0,\\
&& u^{A\la}_m\dr^\mu_A\cL+ u^{A\mu}_m\dr^\la_A\cL=0.
\ee

\section{Multimomentum Hamiltonian formalism}

Let $\Pi$ be the Legendre bundle (\ref{00}) over a fibred manifold
$Y\to X$. It is provided with the fibred coordinates $( x^\la ,y^i,p^\la_i)$:
\[{p'}^\la_i = J \frac{\dr y^j}{\dr{y'}^i} \frac{\dr
{x'}^\la}{\dr x^\m}p^\m_j, \qquad J^{-1}=\det (\frac{\dr {x'}^\la}{\dr
x^\m}). \]
By $J^1\Pi$ is meant the first order jet manifold of
$\Pi\to X$. It is coordinatized by
\[( x^\la ,y^i,p^\la_i,y^i_{(\m)},p^\la_{i\m}).\]

We call by a momentum morphism any bundle morphism
$\Phi:\Pi\to J^1Y$ over $Y$.
Given a momentum morphism $\Phi$, its composition with the
monomorphism (\ref{18})
is represented by the horizontal pullback-valued 1-form
\[\Phi =dx^\la\otimes(\dr_\la +\Phi^i_\la\dr_i)\]
on $\Pi\to Y$. For instance, let $\G$ be a connection on $Y$. Then, the
composition $\wh\G=\G\circ\pi_{\Pi Y}$ is a momentum morphism. Conversely,
every momentum morphism $\Phi$ determines
the associated connection $ \G_\Phi =\Phi\circ\wh 0_\Pi$
on $Y\to X$ where $\wh 0_\Pi$ is the global zero section of $\Pi\to Y$.
Every connection $\G$ on $Y$ gives rise to the connection
\begin{equation}
\wt\G =dx^\la\otimes[\dr_\la +\G^i_\la (y)\dr_i +
(-\dr_j\G^i_\la (y)  p^\m_i-K^\m{}_{\nu\la}(x) p^\nu_j+K^\al{}_{\al\la}(x)
p^\m_j)\dr^j_\m]  \label{404}
\end{equation}
on $\Pi\to X$ where $K$ is a linear symmetric connection  on $T^*X$.

The Legendre manifold $\Pi$ carries the multimomentum Liouville form
\[\th =-p^\la_idy^i\w\om\otimes\dr_\la \]
and the multisymplectic form $\Om$ (\ref{406}).

The Hamiltonian formalism in fibred manifolds is formulated
intrinsically in terms of Hamiltonian connections which play the
role similar to that of Hamiltonian vector fields in the symplectic geometry.

We say that a  connection
$\g$ on the fibred Legendre manifold $\Pi\to X$ is a Hamiltonian
connection if the exterior form $\g\rfloor\Om$  is closed.
An exterior $n$-form $H$ on the
Legendre manifold $\Pi$ is called a  Hamiltonian form if
there exists a Hamiltonian connection  satisfying the equation (\ref{013}).

Let $H$ be a Hamiltonian form. For any exterior horizontal density
$\wt H=\wt{\cH}\om$ on $\Pi\to X$, the form $H-\wt H$ is a Hamiltonian form.
Conversely, if $H$ and $H'$ are  Hamiltonian forms,
their difference $H-H'$ is an exterior horizontal density on $\Pi\to X$.
Thus, Hamiltonian  forms constitute an affine space
modelled on a linear space of the exterior horizontal densities on
$\Pi\to X$.

Let $\G$ be a connection on $Y\to X$ and $\wt\G$ its lift
(\ref{404}) onto $\Pi\to X$. We have the equality
\[\wt\G\rfloor\Om =d(\wh\G\rfloor\th).\]
A glance at this equality shows that $\wt\G$ is a Hamiltonian
connection and
\[ H_\G=\wh\G\rfloor\th =p^\la_i dy^i\w\om_\la -p^\la_i\G^i_\la\om\]
is a Hamiltonian form. It follows that every
Hamiltonian form on $\Pi$ can be
given by the expression (\ref{017}) where $\G$ is some
connection on $Y\to X$.
Moreover, a Hamiltonian form has the canonical splitting (\ref{017})
as follows.
Given a  Hamiltonian form $H$, the vertical tangent morphism
$VH$ yields the momentum morphism
\[ \wh H:\Pi\to J^1Y, \qquad y_\la^i\circ\wh H=\dr^i_\la\cH,\]
and the associated connection $\G_H =\wh H\circ\wh 0$
on $Y$. As a consequence, we have the canonical splitting
\[ H=H_{\G_H}-\wt H.\]

The Hamilton operator $\cE_H$ for a Hamiltonian form $H$
is defined to be the first order differential operator
\begin{equation}
\cE_H=dH-\wh\Om=[(y^i_{(\la)}-\dr^i_\la\cH) dp^\la_i
-(p^\la_{i\la}+\dr_i\cH) dy^i]\w\om, \label{3.9}
\end{equation}
where $\wh\Om$
is the pullback of the multisymplectic form $\Om$ onto $J^1\Pi$.

For any connection $\g$ on $\Pi\to X$, we have
\[\cE_H\circ\g =dH-\g\rfloor\Om.\]
It follows that  $\g$  is a Hamiltonian jet field for a
Hamiltonian form $H$ if and only if it takes its values into
$\Ker\cE_H$, that is, satisfies  the algebraic Hamilton equations
\begin{equation}
\g^i_\la =\dr^i_\la\cH, \qquad \g^\la_{i\la}=-\dr_i\cH. \label{3.10}
\end{equation}

Let a Hamiltonian connection has an integral section $r$ of $\Pi\to X$.
Then, the Hamilton equations (\ref{3.10}) are brought into the first
order differential Hamilton equations (\ref{3.11}).

Now we consider relations between Lagrangian and Hamiltonian
formalisms on fibred manifolds in case of semiregular Lagrangian densities
$L$ when the preimage $\wh L^{-1}(q)$ of each point of
$q\in Q$ is the connected submanifold of $J^1Y$.

Given a Lagrangian density $L$, the vertical tangent morphism $VL$ of
$L$ yields the Legendre morphism
\be &&\wh L : J^1Y\to \Pi, \\
&& p^\la_i\circ\wh L =\pi^\la_i.\ee

We say that a  Hamiltonian form
$H$ is associated with a Lagrangian density $L$ if $H$ satisfies the relations
\bea &&\wh L\circ\wh H\mid_Q=\Id_Q, \qquad Q=\wh L( J^1Y) \label{2.30a},\\
&& H=H_{\wh H}+L\circ\wh H. \label{2.30b}\eea
Note that different  Hamiltonian forms can be associated with the same
Lagrangian density.

Let a  section $r$ of $\Pi\to X$
be a solution of the Hamilton equations (\ref{3.11})
for a Hamiltonian form $H$ associated with a semiregular Lagrangian
density $L$. If $r$ lives on the constraint space $Q$, the section
$\ol s=\wh H\circ r$ of $J^1Y\to X$ satisfies the first
order Euler-Lagrange equations (\ref{306}).
Conversely, given a semiregular Lagrangian density $L$, let
$\ol s$ be a solution of the
first order Euler-Lagrange equations (\ref{306}).
Let $H$ be a Hamiltonian form associated with $L$ so that
\begin{equation}
\wh H\circ \wh L\circ \ol s=\ol s.\label{2.36}
\end{equation}
Then, the section $r=\wh L\circ \ol s$ of $\Pi\to X$ is a solution of the
Hamilton equations (\ref{3.11}) for $H$.
For sections $\ol s$ and $r$, we have the relations
\[\ol s=J^1s, \qquad  s=\pi_{\Pi Y}\circ r\]
where $s$ is a solution of the second order Euler-Lagrange equations
(\ref{2.29}).

We shall say that a family of Hamiltonian forms $H$
associated with a semiregular Lagrangian density $L$ is
complete if, for each solution $\ol s$ of the first order Euler-Lagrange
equations (\ref{306}), there exists
a solution $r$ of the Hamilton equations (\ref{3.11}) for
some  Hamiltonian form $H$ from this family so that
\begin{equation}
r=\wh L\circ\ol s,\qquad  \ol s =\wh H\circ r, \qquad
\ol s= J^1(\pi_{\Pi Y}\circ r). \label{2.37}
\end{equation}
Such a complete family
exists iff, for each solution $\ol s$ of the Euler-Lagrange
equations for $L$, there exists a  Hamiltonian form $H$ from this
family so that the condition (\ref{2.36}) holds.

The most of field models possesses affine and
quadratic Lagrangian densities. Complete
families of Hamiltonian forms associated with such Lagrangian densities
always exist \cite{sard,7sar}.

\section{Hamiltonian gauge theory}

Let us consider the gauge theory of principal connections treated
the gauge potentials.

In the rest of the article, the manifold $X$ is assumed to be
oriented and provided with a nondegenerate fibre metric $g_{\m\nu}$
in the tangent bundle of $X$. We denote $g=\det(g_{\m\nu}).$

Let $P\to X$ be a principal bundle with a structure Lie group $G$
wich acts on $P$ on the right.
There is the 1:1 correspondence between the principal connections $A$ on
$P$  and the global sections of the bundle $C=J^1P/G$. The
finite-dimensional configuration space of principal connections is the
jet manifold $J^1C$ coordinatized by $(x^\mu, k^m_\mu, k^m_{\mu\la}).$

There exists the canonical splitting
\begin{equation}
J^1C=C_+\op\oplus_C C_-=(J^2P/G)\op\oplus_C
(\op\w^2 T^*X\op\otimes_C V^GP), \label{N31}
\end{equation}
\[ k^m_{\mu\la}=\frac12(k^m_{\mu\la}+k^m_{\la\mu}+c^m_{nl}k^n_\la k^l_\mu)
+\frac12( k^m_{\mu\la}-k^m_{\la\mu} -c^m_{nl}k^n_\la k^l_\mu),\]
over $C$. There are the corresponding canonical surjections:
\be &&{\cal S}: J^1 C\to C_+, \qquad {\cal S}^m_{\la\mu}=
k^m_{\mu\la}+k^m_{\la\mu} +c^m_{nl}k^n_\la k^l_\mu,\\
&& \cF: J^1 C\to C_-,\qquad
\cF^m_{\la\mu}= k^m_{\mu\la}-k^m_{\la\mu} -c^m_{nl}k^n_\la k^l_\mu.\ee

The Legendre bundle over the bundle $C$ is
\[\Pi=\op\w^n T^*X\otimes TX\op\otimes_C [C\times\ol C]^*.\]
It is coordinatized by $(x^\mu,k^m_\mu,p^{\mu\la}_m)$.

On the configuration space (\ref{N31}),
the conventional Yang-Mills Lagrangian density $L_{YM}$
is given by the expression
\begin{equation}
L_{YM}=\frac{1}{4\ve^2}a^G_{mn}g^{\la\mu}g^{\bt\nu}\cF^m_{\la
\beta}\cF^n_{\mu\nu}\sqrt{\mid g\mid}\,\om \label{5.1}
\end{equation}
where  $a^G$ is a nondegenerate $G$-invariant metric
in the Lie algebra of $G$. The Legendre morphism
associated with the Lagrangian density (\ref{5.1}) takes the form
\bea &&p^{(\mu\la)}_m\circ\wh L_{YM}=0, \label{5.2a}\\
&&p^{[\mu\la]}_m\circ\wh L_{YM}=\ve^{-2}a^G_{mn}g^{\la\al}g^{\mu\bt}
\cF^n_{\al\bt}\sqrt{\mid g\mid}. \label{5.2b}\eea

Let us consider connections on the bundle $C$ which
take their values into $\Ker\wh L_{YM}$:
\begin{equation}
S:C\to C_+, \qquad
S^m_{\m\la}-S^m_{\la\m}-c^m_{nl}k^n_\la k^l_\m=0. \label{69}
\end{equation}

For all these connections, the Hamiltonian forms
\ben &&H=p^{\mu\la}_mdk^m_\mu\w\om_\la-
p^{\mu\la}_mS_B{}^m_{\mu\la}\om-\wt{\cH}_{YM}\om, \label{5.3}\\
&&\wt{\cH}_{YM}= \frac{\ve^2}{4}a^{mn}_Gg_{\mu\nu}
g_{\la\bt} p^{[\mu\la]}_m p^{[\nu\bt]}_n\mid g\mid ^{-1/2},\nonumber\een
are associated with the Lagrangian density $L_{YM}$ and constitute the
complete family.
Moreover, we can minimize this complete family if we restrict our
consideration to connections (\ref{69}) of the following type.
Given a symmetric linear connection $K$
on the cotangent bundle $T^*X$ of $X$, every principal connection $B$ on
$P$ gives rise to the connection $S_B$ (\ref{69}) such that
\[S_B\circ B={\cal S}\circ J^1B,\]
\[S_B{}^m_{\mu\la}=\frac{1}{2} [c^m_{nl}k^n_\la
k^l_\mu  +\dr_\mu B^m_\la+\dr_\la B^m_\mu -c^m_{nl}
(k^n_\mu B^l_\la+k^n_\la B^l_\mu)] -K^\bt{}_{\mu\la}(B^m_\bt-k^m_\bt).\]

The corresponding Hamilton equations for sections $r$ of $\Pi\to X$ read
\ben &&\dr_\la p^{\mu\la}_m=-c^n_{lm}k^l_\nu
p^{[\mu\nu]}_n+c^n_{ml}B^l_\nu p^{(\mu\nu)}_n
-K^\mu{}_{\la\nu}p^{(\la\nu)}_m, \label{5.5} \\
&&\dr_\la k^m_\mu+ \dr_\mu k^m_\la=2S_B{}^m_{(\mu\la)}\label{5.6}\een
plus the equation (\ref{5.2b}). The
equations (\ref{5.2b}) and (\ref{5.5}) restricted to the constraint space
(\ref{5.2a}) are the familiar
Yang-Mills equations for $A=\pi_{\Pi C}\circ r.$
Different Hamiltonian forms \ref{5.3}) lead to different
equations (\ref{5.6}) which play the role of the gauge-type condition.

In gauge theory, matter fields possessing only internal symmetries are
described by sections of a vector bundle
\[Y=(P\times V)/G\]
associated with a principal bundle $P$. It is provided with a $G$-invariant
fibre metric $a^Y$. Because of the canonical vertical splitting
\[VY=Y\op\times_XY, \]
the metric $a^Y$ is a
fibre metric in the vertical tangent bundle $VY\to X$.
A linear connection $\G$ on $Y$ is assumed to be associated
with a principal connection on $P$. It takes the form (\ref{S4}).

On the configuration space $J^1Y$, the Lagrangian density of
matter fields in the presence of a background connection $\G$ on $Y$
reads
\begin{equation}
L_{(m)}=\frac12a^Y_{ij}[g^{\mu\nu}(y^i_\mu-\G^i_\mu)
(y^j_\nu-\G^j_\nu)-m^2y^iy^j]\sqrt{\mid g\mid}\om.\label{5.12}
\end{equation}

The Legendre bundle of the vector bundle $Y$ is
\[\Pi=\op\wedge^n T^*X\op\otimes_YTX\op\otimes_Y Y^*. \]
The unique Hamiltonian form on $\Pi$ associated with the
Lagrangian density  $L_{(m)}$ (\ref{5.12}) reads
\ben && H_{(m)}=p^\la_idy^i\w\om_\la-p^\la_i
\G^i_\la\om- \frac12(a^{ij}_Yg_{\mu\nu}p^\mu_ip^\nu_j\mid
g\mid^{-1}\nonumber \\
&& \qquad + m^2a^Y_{ij}y^iy^j)\sqrt{\mid g\mid}\om \label{5.13}\een
where $a_Y$ is the fibre metric in $V^*Y$ dual to $a^Y$.
There is the 1:1
correspondence between the solutions of the first order Euler-Lagrange
equations for the regular Lagrangian density (\ref{5.12})
and the solutions of the
Hamilton equations for the Hamiltonian form (\ref{5.13}).

In the case
of unbroken symmetries, the total Lagrangian density of gauge potentials and
matter fields is defined on the configuration space
\[J^1Y\op\times_X J^1C.\]
It is the sum of the Yang-Mills Lagrangian density (\ref{5.1}) and the
Lagrangian density (\ref{5.12}) where
\begin{equation}
\G^i_\la=k^m_\la I_m{}^i{}_jy^j.\label{5.14}
\end{equation}
The associated Hamiltonian forms are the sum of the
Hamiltonian forms (\ref{5.3}) where $S=S_B$ and the Hamiltonian
form (\ref{5.13}) where $\G$ is given by the expression (\ref{5.14}).
In this case, the Hamilton equation (\ref{5.5}) contains
the familiar matter source $p^\mu_iI_m{}^i{}_jy^j$ of gauge potentials.

\section{Energy-momentum conservation laws}

In the framerwork of the multimomentum Hamiltonian formalism, we get the
fundamental identity whose restriction to the Lagrangian constraint
space recovers the familiar energy-momentum conservation law, without
appealing to any symmetry condition.

Let $H$ be a Hamiltonian form on the Legendre bundle $\Pi$
(\ref{00}) over a fibred manifold $Y\to X$. Let $r$ be a section of
of the fibred Legendre manifold $\Pi\to X$ and $(y^i(x), p^\la_i(x))$ its
local components. Given a connection $\G$ on $Y\to X$, we consider the
following $T^*X$-valued $(n-1)$-form on $X$:
\ben &&T_\G(r)=-(\G\rfloor H)\circ r,\nonumber\\
&&T_\G (r)=[p^\la_i(y^i_\mu -\G^i_\mu)-\dl^\la_\mu(p^\al_i(y^i_\al-\G^i_\al)
-\wt{\cH}_\G)] dx^\mu\otimes\om_\la, \label{5.8}\een
where $\wt{\cH}_\G$ is the Hamiltonian density in the splitting
(\ref{017}) of $H$ with respect to the connection $\G$.

Let
\[\tau=\tau^\la\dr_\la\]
be a vector field on $X$. Given a connection $\G$ on $Y\to X$, it
gives rise to the vector field
\[\wt\tau_\G= \tau^\la\dr_\la + \tau^\la\G^i_\la\dr_i +
(-\tau^\m p^\la_j\dr_i\G^j_\m
-p^\la_i\dr_\m\tau^\m + p^\m_i\dr_\m\tau^\la) \dr^i_\la\]
on the Legendre bundle $\Pi$. Let us calculate the Lie derivative
${\bf L}_{\wt\tau_\G}\wt H_\G$ on a section $r$. We have
\begin{equation}
({\bf L}_{\wt\tau_\G}\wt H_\G)\circ r=p^\la_iR^i_{\la\m}+d[\tau^\m
T_\G{}^\la{}_\m (r)\om_\la]-(\wt\tau_{\G V}\rfloor\cE_H)\circ r\label{221}
\end{equation}
where
\be &&R =\frac12 R^i_{\la\m} dx^\la\wedge dx^\m\otimes\dr_i=\\
&&\quad \frac12 (\dr_\la\G^i_\m -\dr_\m\G^i_\la +\G^j_\la\dr_j\G^i_\m
-\G^j_\m\dr_j\G^i_\la) dx^\la\wedge dx^\m\otimes\dr_i; \ee
of  the connection $\G$,  $\cE_H$ is the Hamilton
operator (\ref{3.9}) and
\[\wt\tau_{\G V}=\tau^\la(\G^i_\la-y^i_\la)\dr_i +
(-\tau^\m p^\la_j\dr_i\G^j_\m
-p^\la_i\dr_\m\tau^\m+p^\m_i\dr_\m\tau^\la-\tau^\mu p^{\la\mu}_i) \dr^i_\la\]
is the vertical part of the canonical
horizontal splitting of the vector field $\wt\tau_V$ on $\Pi$
over $J^1\Pi$. If $r$ is a
solution of the Hamilton equations, the equality (\ref{221}) comes
to the identity
\begin{equation}
(\dr_\mu+\G^i_\mu\dr_i-\dr_i\G^j_\mu
p^\la_j\dr^i_ \la)\wt{\cH}_\G-\frac{d}{dx^\la}
T_\G{}^\la{}_\mu (r)= p^\la_iR^i_{\la\mu}. \label{5.27}
\end{equation}
On solutions of the Hamilton equations, the form (\ref{5.8}) reads
\begin{equation}
T_\G(r)=[p^\la_i\dr^i_\mu\wt{\cH}_\G-
\dl^\la_\mu(p^\al_i\dr^i_\al\wt{\cH}_\G-\wt{\cH}_\G)]
dx^\mu\otimes\om_\la.\label{5.26}
\end{equation}
One can verify that the identity (\ref{5.27}) does not depend upon choice
of the connection $\G$.

For instance,
if $X={\bf R}$ and $\G$ is the trivial connection, then
\[
T_\G(r)=\wt{\cH}_0dt
\]
where $\wt{\cH}_0$ is a Hamiltonian and the identity
(\ref{5.27}) consists with the familiar energy conservation law

Unless $n=1$, the identity (\ref{5.27}) can not be regarded directly
as the energy-momentum conservation law. To clarify its physical meaning,
we turn to the Lagrangian formalism.
Let a multimomentum Hamiltonian form $H$ be associated with a
semiregular Lagrangian density $L$. Let $r$ be a solution
of the Hamilton equations for $H$ which lives on the Lagrangian
constraint space $Q$ and $\ol s$ the associated solution  of the
first order
Euler-Lagrange equations for $L$ so that they satisfy the conditions
(\ref{2.37}). Then, we have
\[
T_\G (r)=T_\G(\ol s)
\]
where is the Lagrangian canonical energy-momentum tensor (\ref{S14}).
It follows that the form (\ref{5.26}) may be treated as a Hamiltonian
canonical energy-momentum tensor with respect to a background connection
$\G$ on the fibred manifold $Y\to X$ (or a Hamiltonian
stress-energy-momentum tensor).
At the same time, the examples below will show
that, in field models, the identity (\ref{5.27}) is precisely the
energy-momentum conservation law
for the metric energy-momentum tensor, not the canonical one.

In the Lagrangian formalism, the metric
energy-momentum tensor is defined to be
\[ \sqrt{-g} t_{\al\bt}=2\frac{\dr\cL}{\dr g^{\al\bt}}.\]
In case of a background world metric $g$, this object is well-behaved.
In the framework of the multimomentum Hamiltonian formalism,
one can introduce the similar tensor
\begin{equation}
\sqrt{-g}t_H{}^{\al\bt}=2\frac{\dr\cH}{\dr g_{\al\bt}}.\label{5.28}
\end{equation}

Recall the useful relation
\[
\frac{\dr}{\dr g^{\al\bt}} = -g_{\al\mu}g_{\bt\nu}\frac{dr}{g_{\mu\nu}}.
\]

If a multimomentum Hamiltonian form $H$ is associated
with a semiregular Lagrangian density $L$, we have the equalities
\be
&&
t_H{}^{\al\bt}(q)
=-g^{\al\mu}g^{\bt\nu}t_{\mu\nu}(x^\la,y^i,\dr_\la^i\cH(q)), \\
&&  t_H{}^{\al\bt}(x^\la,y^i,\pi^\la_i(z))=-g^{\al\mu}g^{\bt\nu}t_{\mu\nu}(z)
\ee
where $q\in Q$ and
\[ \wh H\circ\wh L(z)=z.\]
In view of these equalities, we can think of the tensor (\ref{5.28})
restricted to the Lagrangian constraint space $Q$ as being the
Hamiltonian metric energy-momentum tensor.
On $Q$, the tensor (\ref{5.28}) does not depend upon choice of
a Hamiltonian form $H$ associated with $L$. Therefore, we shall
denote it by the common symbol $t$. Set
\[
t^\la{}_\al = g_{\al\nu}t^{\la\nu}.
\]

In the presence of a background world metric
$g$, the identity (\ref{5.27}) takes the form
\begin{equation}
t^\la{}_\al\{^\al{}_{\la\mu}\}\sqrt{-g}
+(\G^i_\mu\dr_i-\dr_i\G^j_\mu p^\la_j\dr^i_
\la)\wt{\cH}_\G=\frac{d}{dx^\la} T_\G{}^\la{}_\mu +p^\la_iR^i_{\la\mu}
\label{5.29} \end{equation}
where
\[\frac{d}{dx^\la} = \dr_\la +\dr_\la y^i\dr_i +\dr_\la p^\m_i\dr_\m^i\]
and by $\{^\al{}_{\la\mu}\}$ are meant the Cristoffel symbols of the
world metric $g$.

For instance, let us examine  matter fields in the presence of a
background gauge potential $A$
which are described by the Hamiltonian form (\ref{5.13}).
In this case, we have the equality
\be && t^\la{}_\mu\sqrt{-g}=T^\la{}_\mu=
[a^{ij}_Yg_{\mu\nu}p^\la_ip^\nu_j(-g)^{-1} \\
&& \qquad -\dl^\la_\mu
\frac12 (a^{ij}_Yg_{\al\nu}p^\al_ip^\nu_j(-g)^{-1}+
m^2a^Y_{ij}y^iy^j)]\sqrt{-g}\ee
and the gauge invariance condition
\[I_m{}^j{}_ip_j^\la\dr_\la^i\wt{\cH}=0.\]
The identity (\ref{5.29}) then takes the form of the energy-momentum
conservation law
\[\sqrt{-g}\nabla_\la t^\la{}_\mu=-p^\la_iF^m_{\la\mu}I_m{}^i{}_jy^j\]
where $\nabla_\la$ is the covariant derivative relative to the
Levi-Civita connection and $F$ is the strength of the background gauge
potential $A$.

Let us consider gauge potentials $A$ described by the complete family
of the multimomentum Hamiltonian forms (\ref{5.3}) where $S=S_B$ and
$K^\bt{}_{\mu\la} =\{^\bt{}_{\mu\la}\}$. On the
solution $k=B,$ the curvature of the connection $S_B$ is reduced to
\be
&& R^m_{\la\al\mu}=\frac12(\dr_\la\cF^m_{\al\mu}-
c^m_{qn}k^q_\la\cF^n_{\al\mu}-\{^\bt{}_{\al\la}\}\cF^m_{\bt\mu}-
\{^\bt{}_{\mu\la}\}\cF^m_{\al\bt})=\\
&& \quad \frac12[(\dr_\al F^m_{\la\mu} - c^m_{qn}k^q_\al F^n_{\la\mu} -
\{^\bt{}_{\la\al}\}F^m_{\mu\bt}) -
(\dr_\mu F^m_{\la\al} - c^m_{qn}k^q_\mu F^n_{\la\al} -
\{^\bt{}_{\la\mu}\}F^m_{\al\bt})].
\ee
Set
\[S^\la{}_\mu=p_m^{[\al\la]}\dr^m_{\al\mu}\wt{\cH}_{YM}=
\frac{\ve^2}{2\sqrt{-g}}a^{mn}_Gg_{\mu\nu}
g_{\al\bt} p^{[\al\la]}_m p^{[\bt\nu]}_n.\]
We have
\[
S^\la{}_\mu = \frac12 p^{[\al\la]}F^m_{\mu\al}, \qquad \wt{\cH}_{YM} =
\frac12 S^\al{}_\al.
\]
In virtue of Eqs.(\ref{5.2a}), (\ref{5.2b}) and (\ref{5.5}), we
obtain the relations
\[p_m^{[\la\al]}R^m_{\la\al\mu}=\dr_\la S^\la{}_\mu
-\{^\bt{}_{\mu\la}\}S^\la{}_\bt,\]
\[\dr^\bt_n\G^m_{\al\mu}p^{\al\la}_m\dr^n_{\bt\la}
\wt{\cH}_{YM}=\{^\bt{}_{\al\mu}\}S^\al{}_\bt,\]
\[t^\la{}_\mu\sqrt{-g}=2S^\la{}_\mu-\frac12\dl^\la_\mu S^\al{}_\al\]
and
\[t^\la{}_\mu\sqrt{-g}=T^\la{}_\mu+S^\la{}_\mu.\]
The identity (\ref{5.29}) then takes the form of the energy-momentum
conservation law
\[\nabla_\la t^\la{}_\mu=0\]
in the presence of a background world metric.

Note the identity (\ref{5.27}) remains true also in gravitation theory.
But its treatment as an energy-momentum conservation law is under discussion.
The key point consists in
the feature of a gravitational field as a Higgs field whose
canonical momenta on the constraint space are equal to zero.

\end{document}